\documentclass{article}
\usepackage[utf8]{inputenc}
\usepackage{authblk}
\usepackage{setspace}
\usepackage[margin=1.25in]{geometry}
\usepackage{graphicx}
\graphicspath{ {./figures/} }
\usepackage{subcaption}
\usepackage{amsmath}
\usepackage{lineno}

\usepackage[style=nejm, 
citestyle=numeric-comp,
sorting=none]{biblatex}
\title{A High Intensity Attosecond Light Source in Compact Geometry at ELI ALPS User Facility}

\author[1$\dag$]{Arjun Nayak}
\author[1$\dag$$\ddagger$]{Mathieu Dumergue}
\author[1*]{Sourin Mukhopadhyay}
\author[1]{Debobrata Rajak}
\author[1,3]{Naveed Ahmed}
\author[1]{J\'anos Csontos}
\author[1]{Szabolcs T\'oth}
\author[1]{Prabhash Prasannan  Geetha}
\author[2$\S$]{Ioannis Orfanos}
\author[2]{Emmanouil Skantzakis}
\author[1,2]{Paraskevas Tzallas}
\author[1,2]{Dimitris Charalambidis}
\author[1]{Katalin Varj\'u}
\author[1]{Subhendu Kahaly}
\author[1*]{Zsolt Diveki}

\affil[1]{ELI ALPS, The Extreme Light Infrastructure ERIC, Wolfgang Sandner u. 3., 6728 Szeged, Hungary}
\affil[2]{Foundation for Research and Technology - Hellas, Institute of Electronic Structure \& Laser, PO Box 1527, GR71110 Heraklion (Crete), Greece}
\affil[3]{Institute of Physics, University of Szeged, H-6720 Szeged, Hungary}
\affil[*]{Address correspondence to: sourin.mukhopadhyay@eli-alps.hu, zsolt.diveki@eli-alps.hu}
\affil[$\dag$]{These authors contributed equally to this work.}
\affil[$\ddagger$]{Current affiliation: Laboratoire pour l’Utilisation des Lasers Intenses, CNRS-Ecole Polytechnique-CEA-Sorbonne Universit\'e}
\affil[$\S$]{Current affiliation: School of Engineering, Mediterranean College – University of Derby, Athens, Greece}

\date{}

\begin{document}

\maketitle

\begin{abstract}
High-order harmonic generation (HHG) has become a standard technique for producing attosecond XUV pulses in the laboratory, yet the high flux necessary for nonlinear XUV photoionization remains accessible to only a few research groups. Here, we introduce the SYLOS Compact high-harmonic beamline at ELI ALPS, specifically designed to provide the flux required for non-linear optics in the XUV. We present a detailed characterization of the beam line demonstrating its capability to generate and utilize both intense attosecond pulse trains and isolated attosecond pulses. We further showcase the two-XUV-photon double ionization of neon (Ne) and argon (Ar), achieved in a user campaign. The results underscore the beamline’s capability to support cutting-edge attosecond experiments and investigations of ultrafast electron dynamics on the attosecond scale.
\end{abstract}


\section{Introduction}
The invention of laser based sources of coherent XUV and soft X-ray radiation about three decades ago has spurred tremendous advancement in the field of Atomic Molecular and Optical (AMO) sciences \cite{Krausz2009} facilitating an ever clearer understanding of processes in the electronic system of atoms and small molecules \cite{Nisoli2017}, liquid \cite{Longetti2020} and solid phase targets \cite{Park2021}, while opening up the possibility to extend nonlinear optics to the XUV domain \cite{Orfanos2020}.
Not only did the novel technology bring tools that were until now only available at national facilities such as synchrotrons and FELs to the hands of a medium-sized laser laboratory, but it also lead to the birth of an entirely new field in physics, namely attosecond sciences \cite{Reduzzi2015}.

XUV radiation generated by the process of High-Order Harmonic Generation in Gasses (GHHG) offers excellent intrinsic temporal and spatial coherence properties, which culminate in the creation of attosecond pulses that can be focused down to sub-micrometer dimensions \cite{Major2020}.
The fact that the attosecond pulses are also phase-locked to the fundamental driving field is exploited routinely to carry out pump-probe experiments that can address the intrinsic timescale associated with the electron dynamics in various phases of matter. 
The advancement of femtosecond laser technology and the continuous progress in R\&D in GHHG systems, have enabled this unique tool to become almost a mature technology in multiple laboratories around the world. However, technical constraints still require source specialization and optimization in terms of XUV pulse energy, repetition rate and photon energy. This task is far from straightforward, as these properties are inherently tied to the driving laser's parameters, including beam profile, polarization, and wavefront quality. They also depend on the design and operational parameters of the beamline, which must follow specific scaling laws \cite{Rudawski2013, Heyl2016}. Consequently, achieving the desired performance requires a demanding laser system and a sophisticated, customized beamline design.

The capability to generate energetic XUV pulses at suitable repetition rates using GHHG opens promising directions exploring atomic and molecular systems. One important pathway is the exploration of nonlinear XUV science \cite{Orfanos2020}, where multi-photon processes have already been observed with GHHG sources \cite{Nayak2018, Manschwetus2016, Major2021}, and XUV–XUV pump–probe schemes have been successfully demonstrated \cite{Tzallas2011, Takahashi2013}. Building on these achievements, GHHG further enables the investigation of attosecond electron and nuclear dynamics under conditions of minimal dressing-field \cite{Palacios2014}, offering a more direct probe of ultrafast correlated electron dynamics. The relevance of this approach lies in overcoming the limitations of conventional XUV–IR pump–probe schemes, which require moderate to strong fields on the order of $10^{11}$ W/cm$^{2}$ that inevitably perturb the system and complicate theoretical and experimental analysis. In contrast, XUV–photon interactions generally occur within the perturbative regime, thereby offering direct insight into the intrinsic dynamics of the system under study. \cite{Palacios2009, Palacios2014a}. Furthermore, the strong nonlinearity of GHHG enables the generation of pulses substantially shorter than the driving field, allowing inner-shell and core excitations to be probed on timescales preceding relaxation processes \cite{Bergues2018}. They provide a comprehensive view of electron–electron interactions that govern the dynamics of the system immediately following the creation of a hole state.
Going one step further, through the application of gating techniques \cite{Mashiko2008} and pulse compression \cite{Sansone2006}, single XUV pulses can be isolated from the pulse train, bringing the time resolution approaching the atomic unit of time $\approx$ 24~\text{as} \cite{Gaumnitz2017}. A prominent example for the application of such pulses is the process of two-photon double ionization which explores the correlations of electrons occurring on different time scales in the femtosecond to attosecond regime in multidimensional space \cite{Palacios2010} and to which GHHG with high pulse energy and high repetition rate currently seems to be the only viable approach \cite{Hasegawa2005, Antoine2008}.

To date, only a few attosecond pump-probe spectroscopy experiments have been conducted and currently only two, very different methods lead to such experiments. The first, relies on the usage of FEL sources. For a long time they have been the ideal tools to achieve non-linear XUV/X-ray photoionization \cite{Ullrich2012}, particularly with their recent developments on attosecond pulse generation and application \cite{Duris2019, Guo2024, Li2024}. The realization of pump-probe measurements with attosecond resolution with these systems require in depth knowledge of FEL physics. The second, is the tabletop GHHG-based attosecond pump-probe technique. To improve the photon flux and realize isolated attosecond pulse generation, most of the groups used high energy lasers (> 50 mJ) at moderate repetition rate (10-100 Hz), with loose focusing (several meters \cite{Nayak2018, Kretschmar2022}) and various techniques to produce near-single-cycle laser pulses, like polarization gating \cite{Tzallas2011}, hollow-core fiber compression \cite{Kretschmar2024, Sobolev2024} or waveform synthesizers \cite{Takahashi2013, Bergues2018}. Furthermore, several solutions  were proposed to achieve optimal phase matching in the target to increase the harmonic flux \cite{Hareli2020}. The simple scenario of using single gas jet or gas cell was proved both theoretically \cite{Weissenbilder2022} and experimentally \cite{Appi2023} to result in similar harmonic yields. The simplest technical improvement to increase harmonics flux through quasi-phase matching is the usage of target density modulation via changing the distance between multiple gas jets \cite{Seres2007, Pirri2008}. From these it is easy to see that one has to take into account many aspects to design an attosecond beamline dedicated to non-linear XUV attosecond pump-probe studies.

In this communication we present the implementation of ELI ALPS's \cite{Kuhn2017, Shirozhan2024} newly commissioned SYLOS Compact GHHG beamline which is driven by the few-TW class SYLOS (1kHz) or by the SEA (10 Hz) laser systems and can deliver XUV pulses in the $\mu$J range as well as high photon energies reaching up to 150 eV generating both from (multiple) gas jets or from gas cells.
We also confirm the attosecond duration of pulses in a few femtosecond pulse train through the RABBITT technique.
The capabilities of the beamline with respect to the study of non-linear XUV processes are demonstrated by the results of a non-linear interaction of XUV pulses with noble gas atoms using the integrated diagnostic and experimental end stations.
Thereby we introduce to the user community a novel and unique experimental tool which can concurrently deliver XUV attosecond pulses with high pulse energies, high photon energies and a repetition rate compatible with coincidence detection schemes.

\section{Materials and Methods}
\begin{figure*}
    \centering
    \includegraphics[width=15cm]{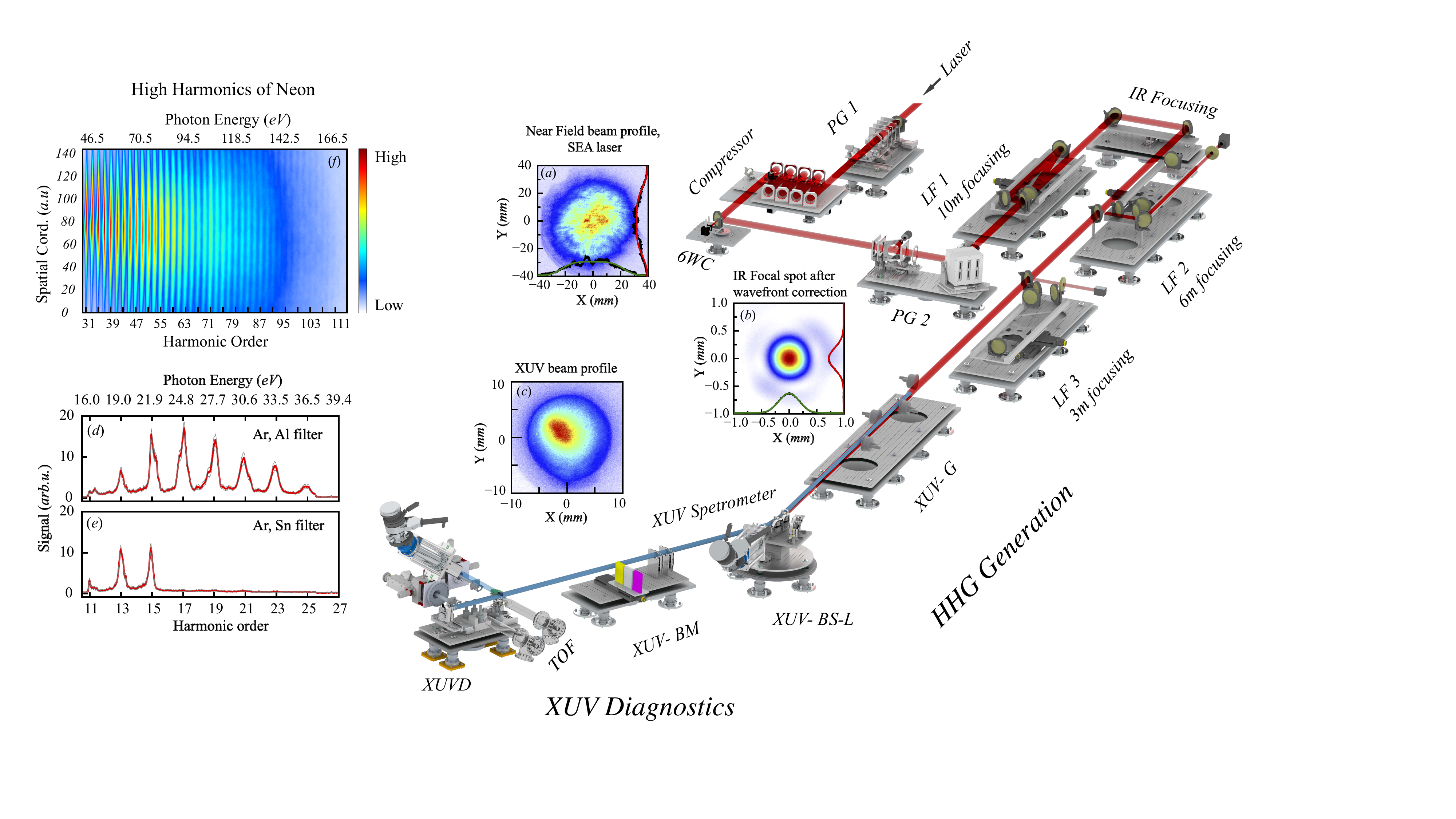}
    \caption{ A full schematic of the Compact Beamline at ELI ALPS depicting the individual sections of beam propagation, associated diagnostics and results. PG1 chamber for transmissive optics, Compressor chamber with a set of chirped mirrors, 6WC for beam steering, Deformable mirror (DM) positioned at $45^\circ$ in PG2, Focusing section consisting of three chambers LF1-3 with f=10 m, 6 m and 3 m focusing mirrors, and the turning chambers IR-St. XUV-G hosts gas targets in the form of gascell and gasjets. XUV-BS-L, XUV-BM, XUVD are for beam separation, beam management and diagnostics respectively (\textit{a}) The near field profile of the SEA laser, (\textit{b}) Measured focal spot after wavefront correction, (\textit{c}) XUV beamprofile, XUV spectra generated in Xenon with (\textit{d}) Al filter  and (\textit{e}) Sn filter on the beampath. (\textit{f}) A typical high harmonic spectra of Neon.}
    \label{fig_full_beamline}
\end{figure*}

The fundamental beam for the Compact beamline is primarily provided by the Single Cycle Laser System (SYLOS2, manufactured by EKSPLA Ltd. and Light Conversion) \cite{Toth2020}. 
The system is a passively CEP-stabilized NOPCPA system, driven by the second harmonic of diode pumped Nd:YAG amplifiers. 
At the output, the system achieves a peak power over 4.5 TW corresponding to 35 mJ at 1 kHz repetition rate. 
The spectrum is centered around 850 nm and it is not fully compressed in time before the beamline.
The pre-compressed pulse 
propagates 35 m under high vacuum conditions to the Compact beamline's PG1 chamber, which hosts transmissive optics, and then  it reaches full compression (SYLOS2 $\approx$\: 12fs) after passing through a set chirped mirror compressor (PC1705, UFI, nominal GDD $\approx$\:  100 fs$^{2}$) just before focusing (see Fig.\ref{fig_full_beamline}).
An acousto-optic programmable dispersive filter (AOPDF, DAZZLER, FastLite Ltd.) permits the fine-tuning of the spectral phase and amplitude to obtain optimal conditions for harmonic generation on the target.
The wavefront and stability are optimized by a  deformable mirror and a pointing stabilizing system.
Consequently, a Strehl ratio of $>$ 0.9 and pointing fluctuations $< 2$ $\mu$rad are achieved, which are essential for efficient and stable harmonic generation in the beamline.

A secondary laser system can be used to align and optimize the beamline, and  also for experiments that do not require a high repetition rate. 
This is the SYLOS Experiment Alignment laser (SEA) which provides up to 35 mJ in 12 fs pulses centered around 850 nm at a repetition rate of 10 Hz. In addition to the control parameters of the SYLOS2 laser, SEA provides easy control over the pulse energy since the thermal loads are significantly lower than for the 35 W average power of the SYLOS2 laser.

To drive the process of GHHG efficiently, the laser pulses must be focused to peak intensities on the order of $10^{13-15}$~W cm$^{-2}$ - depending on the generating gas species - and the area of the interaction must be optimized for phase matching \cite{Heyl2016}. 
For TW class driving lasers, this dictates a loose focusing geometry with an f-number (f$\#$) on the order of 300-500. In the Compact beamline this is accomplished by three independent all-reflective focusing stages (see LF1, LF2 and LF3 respectively in Fig.\ref{fig_full_beamline}) which can be moved into the beam without changing the beam propagation. 
The focal distances are 10 m, 6 m and 3 m corresponding to f$\#$'s of 200, 120 and 60 assuming a full beam with a diameter of 50 mm.
The 10 m focusing stage has been augmented with a 2:1 all-reflective telescope resulting in an effective focal length of 20 m with an increased f$\#$ of 400. As the spherical optics are used at oblique incidence, aberrations cannot be ruled out despite the effort of reducing all angles of incidence to a few degrees. 
We use a deformable mirror placed just before the focusing stages (PG2 chamber) to pre-compensate the resulting wavefront errors. The wavefront is corrected using a wavefront sensor placed in a diagnostic arm located next to the beamline. 
By closing the loop for the wavefront optimization, we can achieve a nearly perfect Gaussian shaped spot around the target position. Under typical generation conditions, the effective beam diameter is reduced to about 40 mm using a hard aperture iris, truncating the super-Gaussian beam of the laser at about the FWHM diameter and leading to the slight halo around the central spot. 
Special care was taken to improve both the intensity distribution and wavefront of the fundamental beam in the interaction region for both have a direct impact on the generation efficiency and XUV beam quality \cite{Dacasa2019, Bandulet2008, Villoresi2004}. An example of the laser beam's profile and the resulting intensity distribution in the focal plane is shown in Fig. \ref{fig_full_beamline}(a) and (b). 

Given the laser pulse parameters and taking into account the losses on mirrors, approximately 30 mJ reaches the gas target with 12 fs pulse duration in a focal spot size of 370 $\mu$m ($e^{-2}$ radius) which results in an estimated peak intensity on the order of $10^{14}$ W.cm$^{-2}$. 
For Xenon as the generating medium, a single atom cutoff according to $E_{ph}=I_p + 3.17 U_p$ of $E_{ph}= $ 47 eV is estimated \cite{Chang2011}.

Generation is accomplished in the XUV-G chamber which is connected to the focusing section through a differential pumping stage.
In this way, the pressure in neighbouring sections is kept below $10^{-6}$ mbar even when the pressure in the generation chamber rises into the $10^{-3}$ mbar region during operation of the gas targets at 1 kHz.
Two different types of gas targets are available for generation: pulsed gas jets or static gas cells.
These two types of targets differ mostly in their phase matching mechanism.
While the single particle response is determined by the properties of the fundamental field and the atomic or molecular species \cite{Lewenstein1994, Nayak2018, Sansone2004}, the macroscopic response of the target depends significantly on the constructive buildup of the XUV field due to phase matching \cite{Rudawski2013, Salieres1999,  Popmintchev2009, Major2019}.
Consequently, achieving phase matching has been a major direction of past research efforts.
Geometrical parameters like focusing, target length, density and position, as well as dynamical parameters like space- and time-dependent ionization and propagation need to be considered.
Depending on the type of target, an extended cell or a localized jet, phase matching is dominated by slightly different contributions \cite{Rudawski2013}. 
However, experimental results show  that both target types can achieve comparable peak conversion efficiencies \cite{Weissenbilder2022, Appi2023, Rivas2018, Takahashi2003a, Hergott2002, Nayak2018, Makos2020} and so it often is the simplicity of a cell or the reduced gas load of the pulsed jet that favours one over the other. 
At 1 kHz repetition rate, the latter advantage is lost. 
The Compact beamline applies both types of targets, and preliminary results indicate that the XUV beam generated from a cell has a slightly better spatial quality than that from pulsed jets (see in Section 3.3). 
Up to four gas jets can be placed in series to enhance the conversion yield through quasi phase matching \cite{Nayak2018}, though in practice the greatest benefit is obtained with only two jets. Optimization of the XUV yield with Argon in a 30 cm long cell with 1.5 mm diameter pinholes requires a  typical cell pressures of 1 mbar, which raises the chamber pressure to $10^{-4}$ mbar. 
Using pulsed gas jets (APCV3, manufactured by MassSpecpecD BV) with 4 bar of backing pressure results in similar chamber pressures at 10 Hz operation, but in a few $10^{-3}$ mbar at 1 kHz operation.
The particle density at the output of the nozzle has been measured using interferometric methods \cite{Nagyilles2023}.
\begin{figure}
    \centering
    \includegraphics[width=8cm]{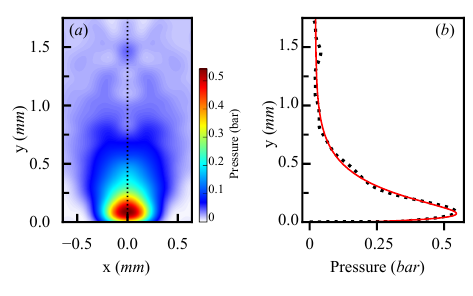}
    \caption{(\textit{a}) Radial dependence of the gas density distribution for the expansion of argon into vacuum from an APCV3 nozzle with 500 $\mu m$ orifice diameter, and peak density along the nozzle axis. Assuming a temperature of 300 K, the corresponding peak pressure would reach 0.5 bar for a density of $1 \times 10^{25}\, m^{-3}$.
    (\textit{b}) Pressure profile taken at the center of the nozzle, black dotted line on (\textit{a}). Red curve: Fit of the  profile by an exponential decay, yielding a decay length of 240 $\mu$m. } 
    \label{Pressure_profile}
\end{figure}
The most characteristic feature is the rapid drop of particle density along the axis of the nozzle following an exponential function with decay length of about 240 $\mu$m (Fig. \ref{Pressure_profile}). 
Assuming that the laser focus is located slightly more than one waist from the nozzle around 0.5 mm, the corresponding peak pressure is around 200 mbar.
However, there is also an appreciable gradient along the expansion axis ranging from 500 mbar at the inner to 150 mbar at the outer edge of the focal spot, which may influence the quality of the XUV beam.
Therefore, we also have applied a custom built gas jet with a slit shaped nozzle of 0.3 x 3 mm, which provides more homogeneous density distribution in the generation region owing to a quasi two-dimensional expansion geometry. The harmonic yield is less sensitive to the precise jet positions along the axis of the jet.  

Following generation, the co-propagating IR and XUV beams have three pathways (upper, middle and lower; the last two paths are not shown in Fig. \ref{fig_full_beamline} for the sake of simplicity) which can be selected in the beam separation chamber (BSL chamber). These paths differ in the way the unconverted IR beam is suppressed which also has implications on the spectral transmission of the XUV beam.

A straight path (the middle one) currently leads directly to a slit-based flat field  XUV spectrometer (Ultrafast Innovations GmbH). The XUV beam isolation is accomplished by a high power aperture with a diameter of 5 mm located 2.5 m from the focus, exploiting the difference in beam divergence between the XUV and IR beams. 
Further suppression is provided by an adjustable iris and optional filter foils that are located about 5 m from the focus. A visible-blind XUV photodiode with an integrated aluminum bandpass filter (Optodiode AXUV100Al) can be inserted into the beam path to measure the mean XUV pulse energy. 
The straight path provides the highest XUV flux but also requires sufficient distance for beam expansion to prevent damage to the foil filters. In the present geometry, aluminum filters can withstand the power of the 1 kHz beam but the Sn filters suffer destruction.

The upper path (Fig. \ref{fig_full_beamline}) can be selected through the reflection from a polished silicon plate, arranged at Brewster angle (AOI $75^\circ$) for the IR beam, and thus providing significant suppression up to 3 OD for the horizontally polarized fundamental beam. 
The beam then enters a beam diagnostic chamber which contains various filter foils, two calibrated XUV photodiodes and an XUV beam profiler based on a one-inch microchannel plate/phosphor combination. 
One of the phododiodes is identical to that in the straight path so that the reflectivity of the silicon plate ($R_{Si,XUV}$) can be determined by comparing their signals. 
We obtain $R_{Si,XUV} \approx 40 \%$ for the lower harmonics in the Sn filter bandpass around 20 eV. 
It should also be noted, that the spectral reflectivity for uncoated silicon exhibits a cutoff around 70 eV according to tabulated data \cite{Henke1993} providing a source for experiments up to this photon energy. 
Moreover, an oxide layer and finite surface roughness are expected to reduce the reflectivity from its' theoretical value \cite{Midorikawa2008}.
Other spectral regions can be accessed by applying thin coatings to the mirror, e.g. a rhodium or silver coating would result in good reflectivity around 120 eV, while NbN enhances the 70 eV range.

The XUV beam passes through a differential pumping stage into the experimental chamber XUV-D. The equipment used in this report consists of a slit-less flat-field XUV spectrometer (Ultrafast Innovations GmbH) and a bipolar ion/electron time-of-flight spectometer (ToF, ETF15, Stefan Kaesdorf GmbH). 
The XUV spectrometer is mounted on top of the chamber and receives the XUV beam after a grazing incidence reflection from another silicon plate. The repeller of the ToF has been replaced by a pulsed gas jet that has been adapted to floating ground operation (APCV3, floating ground). 
In this way, the target gas can be supplied directly to the experimental interaction region while at the same time establishing a potential of 3 kV with respect to the ToF for efficient ion collection and the suppression of space charge effects. 
The XUV beam was focused into the target region in a back-focusing geometry using a multilayer mirror with 50 mm focal length designed for a central wavelength of 40 eV, a bandwidth of 5 eV FWHM and a peak reflectivity of $40 \%$ (Fraunhofer-Institut IOF). This XUV focusing mirror can be replaced with other mirrors reflecting different  central wavelengths.

To make the spectral characterization of the harmonics complete, the setup can perform RABBITT measurements using the aforementioned ToF tube with the cross-correlated XUV-IR beams to retrieve the temporal profile of the attosecond pulse train. The IR beam used for generating the harmonics and the IR used in the RABBITT measurements are co-propagating along the whole beamline. The time delay between them is introduced by rotating a small thin disk placed in the common beampath. 

The lower arm is an optical replica of the upper arm and is configured for specialized experimental implementations or custom user endstations. It interfaces with the second beam diagnostic chamber via a differential pumping stage, achieving pressures below $10^{-8}$mbar, comparable to those in the upper arm.
Due to their optical equivalence, XUV pulses thoroughly characterized in the upper arm can be directed to the lower arm without alteration, enabling experimental applications with well-defined pulse parameters and facilitating precise interpretation of the results.

Other equipment installed in the experimental chamber includes an ion microscope (Stefan Kaesdorf GmbH) recording the focus of a grazing incidence focusing mirror of the Wolter type (Thales SESO SAS) with a focal length of 18.7 cm. 
It can be used for spatially resolved photo-generated ions to characterize the XUV pulses and obtain non-linear optical properties of the target particles \cite{Bergues2018, Tzallas2018}. 
Another important device in connection with the bipolar ToF is a split-and-delay unit which is essential for the major target application of the Compact beamline, namely XUV-XUV pump-probe experiments. This device holds a pair of D-shaped XUV mirrors which produce two overlapping XUV foci in the target region of the ToF by wavefront splitting. 
The delay between the pulses is controlled by a piezoelectric stage with a stepsize of 2 nm. Several wavelength ranges (corresponding to photon energies of $<$20, 22, 40 and 60 eV) and two focal distances (5 and 15 cm) are available.

\section{Results and Discussion}
\subsection{XUV spectrum}

\begin{figure}
    \centering
    \includegraphics[width=8cm]{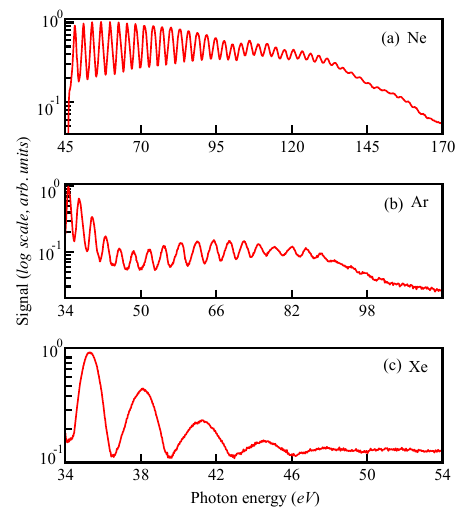}
    \caption{XUV spectra generated by the SEA laser using the pinhole shaped nozzle gas jet (GJ2) with (\textit{a}) Neon,  (\textit{b}) Argon and  (\textit{c}) Xenon.}
    \label{XUV_spectra_center}
\end{figure}
XUV pulses were produced with both the SEA and the SYLOS lasers and a variety of targets. In Fig. \ref{XUV_spectra_center} we show XUV spectra for (a) Neon, (b) Argon and (c) Xenon as the generating gas to present the range of photon energies that can be obtained in the beamline. The spectrometer was calibrated using the known absorption edge of aluminum at 72 eV. 
Neon, having the highest $I_p$ and the highest ionization saturation intensity, reaches a cutoff around 150 eV, while for Xenon the cutoff harmonics are around 45 eV. The spectrum of Argon is more modulated due to the Cooper minimum at 52 eV but still reaches a cutoff around 90 eV. 
The spectrometer in the straight path was used without filters to avoid spectral distortions. In the slit-based configuration, the spectral range has a lower limit of 30 eV and thus lower harmonics cannot be observed. These are presented for Xenon in the upper spectrum  in Fig. \ref{fig_full_beamline} (d,e) to include the harmonic orders 13 and 15 (at 19 and 22 eV) which fall within the transmission range of the bandpass filter used.

\subsection{XUV pulse energy}

To obtain the XUV pulse energy we use an uncoated XUV silicon photodiode (AXUV576, Optodiode) as the spectral response of the solar-blind version may change over time due to the aging of the aluminum overcoat. 
This means that the transmission of the filter used to reject the IR light and the signal from the residual IR light must be measured accurately. 
To this end, one must compare the signals from two Sn filters in series to that of each individual filter and subtracting the residual IR background measured after the insertion of a glass plate to block any XUV light in the beam path for each measurement. 
We obtain for the 200 nm thick Sn filter transmissions of 4.4$\%$ at harmonic order (HO) 13 and 7.8$\%$ at HO 15, averaging to about 6$\%$ in the Sn transmission window. Using tabulated data \cite{Henke1993} and assuming an oxide layer of 4 nm SnO$_2$ on each side of a foil of 200 nm total thickness, a transmission of 10-12$\%$ is estimated, which is almost twice as high as that measured.
The discrepancy may arise from the underestimation of the oxide layer's thickness \cite{Puppin2019, Hurwitz1985} or the presence of carbon contamination on the filter \cite{Hemmers2012}. 
We use the measured filter transmission and the reflectivity of the XUV silicon mirror to estimate the XUV pulse energy at generation. For a value of 7.4 nJ on the photodiode generated from a 30 cm long gas cell filled with xenon and using the SEA laser, we obtain 308 nJ for two harmonics. Assuming an average of six harmonics contributing to the XUV spectrum in the range of 15-40 eV, a total generated pulse energy of 1 $\mu$J can be estimated (about $2 \times 10^{11}$ photons). 
Similar results were obtained also for the SYLOS laser and using a single GJ2 gas jet delivering a pulse energy of 3.8 nJ to the photodiode or 158 nJ at generation for two harmonics.

The pulse energy can be increased by accomplishing quasi phase matching through a series of gas jets positioned along the laser focus. In the present case, two different gas jets were combined, a home-built jet with a slit-shaped nozzle GJ1 (built at FORTH-IESL) and an APCV3 pinhole shaped nozzle of 500 $u$m diameter - GJ2. Each individual target produced an XUV pulse energy of about 4.7 nJ (GJ1) and 4.2 nJ (GJ2) when driven with the SYLOS laser. But when operated simultaneously, the energy for two harmonics in the Sn window reached 7.1 nJ corresponding to 300 nJ at the source ($\approx 1\mu$J for six harmonics).

\subsection{The XUV beam's spatial profile}

\begin{figure}
    \centering
    \includegraphics[width=8cm]{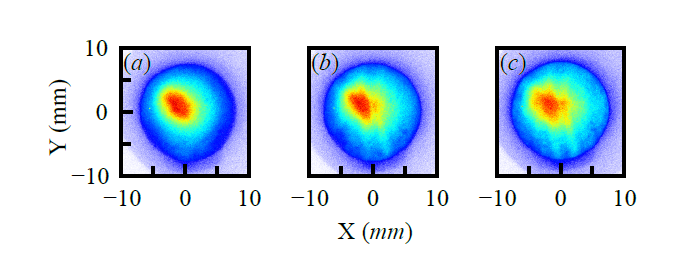}
    \caption{XUV beam profiles (\textit{a}) for the GJ1 gas jet, (\textit{b}) for the GJ2 gas jet and (\textit{c}) both together. The spot size diameter is about 8\,mm and does not change with the number of sources.}
    \label{XUV_profiles}
\end{figure}

For non-linear experiments using these XUV pulses and thus achieving high peak intensities, two other parameters are of importance: focusability and pulse duration. We show in Figure \ref{XUV_profiles} the XUV beam profiles for the two gas jets and their combination.
The XUV beams exhibit a distribution close to a Gaussian with a spot size diameter around 8 mm. For a propagation distance of 6 m from the target this would correspond to a full divergence angle of 1.3 mrad which is about half that of the IR beam.
It can be estimated that the beam size at the XUV back-focusing mirror located 7.5 m from the target would approach 10 mm. 
While the profile for the  GJ1 is smooth, the profile for the GJ2 jet shows some structure in the vertical direction which may arise from wavefront nonuniformity caused by the target density gradient in that direction \cite{Nagyilles2023}. The perturbation persists when the two jets are combined with the GJ2 jet placed behind the GJ1 jet. 

\subsection{Temporal characterization of the XUV pulse}

The metrology of attosecond pulses is still an actively researched area \cite{Orfanos2019} especially in the emerging few-cycle pulse regime, where simplifying assumptions - such as infinite and invariable pulse trains or the existence of only a single trajectory - lose their validity and pulse fluctuations come into play. 
Investigations of the attosecond pulse duration generated with similar laser pulse parameters as ours show that careless application of a retrieval algorithm that does not account for varying IR pulse parameters will lead to an underestimation of both the number of attosecond pulses as well as their individual duration \cite{Osolodkov2020, Mikaelsson2021}. 
Better results are found with more elaborate algorithms that take into account the few-pulse nature as well as fluctuations of driving and harmonic pulse parameters. 
In this sense, our measurement should be considered as a first estimate of the attosecond pulse duration and as a proof-of-principle experiment for the beamline technology.

With a driving IR pulse duration of 12 fs—within the few-cycle regime—the generation of an attosecond pulse train comprising roughly 8 sub-cycle bursts is expected at lower photon energies. This expectation is consistent with the harmonic comb structure observed in the XUV spectrum (Fig. \ref{XUV_spectra_center}). To probe the temporal characteristics of this emission, we employ the RABBITT (Reconstruction of Attosecond Beating by Interference of Two-Photon Transitions) technique. This is an interferometric approach that cross-correlates an XUV comb of high-order harmonics with a phase-locked, time-delayed IR probe derived from the driving field. By recording photoelectron spectra produced via two-photon, two-colour ionization, the method provides access to both the temporal envelope and the spectral phase of the XUV emission.

The Compact beamline incorporates a concentric split-and-delay unit \cite{Paul2001, Kruse2010} within the PG1 chamber, consisting of two fused silica plates with non-overlapping optical paths: an inner probe beam and an outer generation beam. A relative temporal delay on the order of a few femtoseconds is introduced by tilting one of the dispersive plates, thereby modifying the group delay through material dispersion. The relative timing between the IR and XUV pulses is controlled by this delay unit; however, due to its design, only positive delays between the IR and XUV pulses are accessible, ensuring that the XUV pulse always precedes or coincides with the IR pulse within the available delay range. Following the interaction region, the two beams propagate collinearly and are subsequently separated spatially by means of a ring-shaped iris in combination with a high-power pinhole positioned at conjugate image planes before and after the focusing mirror. Owing to its reduced divergence, the XUV beam predominantly transmits through the pinhole. The generated high-order harmonics, together with a residual fraction of the fundamental IR field, are then refocused by a gold mirror and recombined at the second gas jet, located in the detection region of a time-of-flight (ToF) spectrometer for photoelectron spectroscopy. Ionization of atoms by the XUV field in the presence of the IR field initiates a two-photon transition process, giving rise to sideband formation due to constructive interference between competing quantum pathways.
\begin{figure}
    \centering
    \includegraphics[width=8cm]{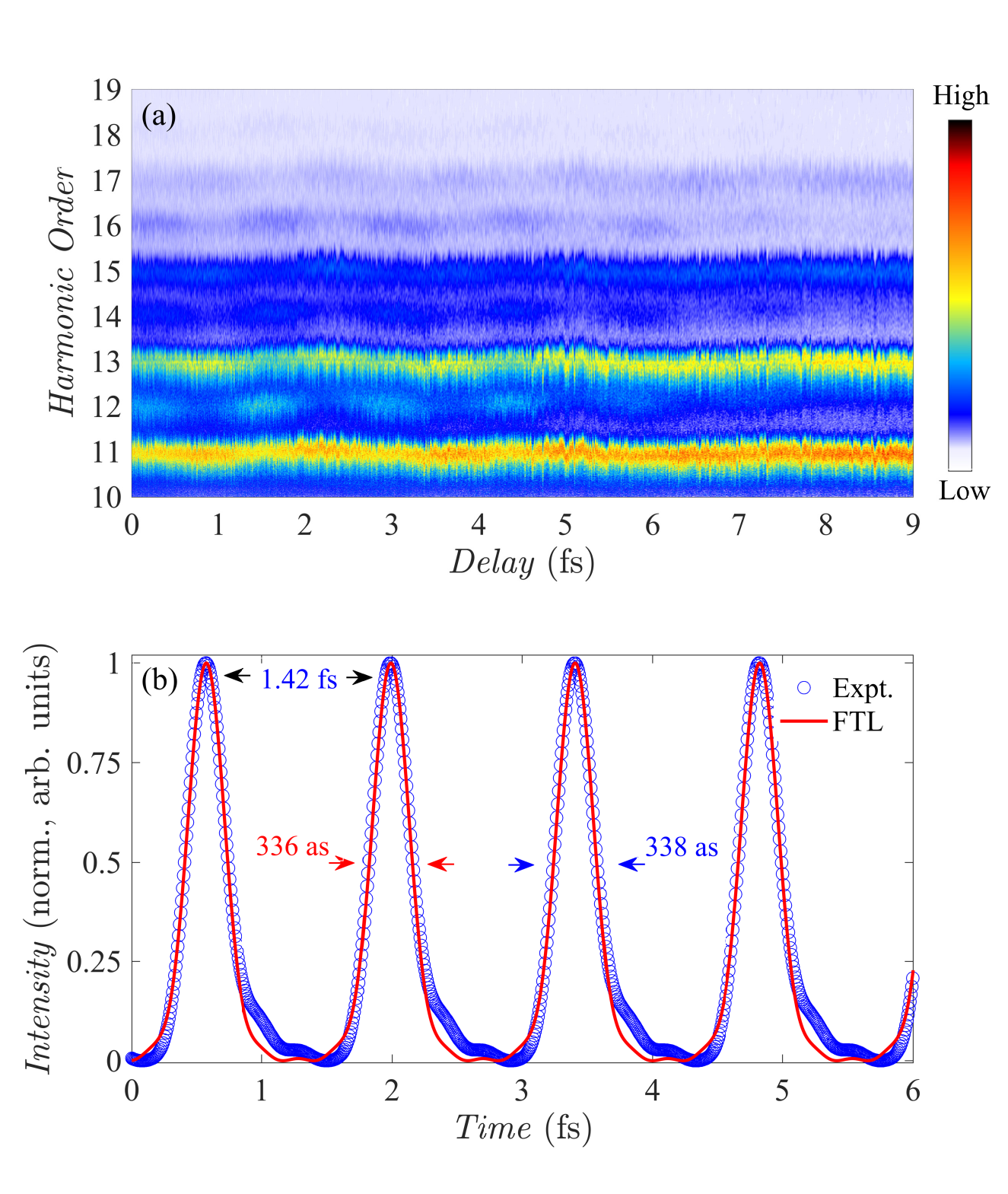}
    \caption{RABBITT trace (\textit{a}) Evolution of the photoelectron spectra generated in argon as a function of the delay plate angle. (\textit{b}) Retrieved attosecond pulse duration. The blue one shows reconstructed pulse from the photoelectron spectra, the red curve is the Fourier transform limitted one.  }
    \label{Rabbitt_map}
\end{figure}
Spectra acquired for XUV pulses generated in argon using the SYLOS laser reveal distinct sideband oscillations at small plate tilt angles (Fig.\ref{Rabbitt_map} \textit{a}).  
The specified thickness mismatch between the plates is less than 1~\(\mu\)m, implying a maximum group delay mismatch of approximately 1.3~fs. As the relative delay increases, the sideband amplitude gradually decreases and vanishes around 7~fs, consistent with the temporal overlap of the two IR pulses. From this, the sideband envelope exhibits a FWHM of 7.5~fs. Assuming Gaussian temporal profiles and an IR pulse duration of 7~fs, the duration of the XUV pulse is estimated to be 2.8~fs. 

These oscillations exhibit a periodicity of \( 2\omega_{\mathrm{IR}} \), consistent with two-photon interference. The phase of the sidebands encodes the relative spectral phase between adjacent harmonics, enabling complete temporal reconstruction when combined with the measured spectral amplitudes. Following (Fig. \ref{Rabbitt_map}\textit{a}), the temporal profile of the XUV pulse train is reconstructed through Fourier transform, given by:

\begin{equation}
I(t) = \left| \sum_q A_q e^{-iq \omega_{\mathrm{IR}} t+ i \phi_q} \right|^2
\label{eq:hh_sum}
\end{equation}
 Here, $A_q$ represents the amplitude of each harmonic and $\phi_q$ is the harmonic phase, it is the phase of harmonic $q$ averaged over the entire pulse train. Using harmonic orders $q = [11, 13, 15, 17, 19 ]$, associated spectral phases $\phi_q = [0, 2.85, 5.23,7.51, 10.15]$~rad and spectral intensities $A_q^2 = [3.02, 2.73, 1.71, 0.79,0.33]$, the reconstructed intensity envelope yields (Fig. \ref{Rabbitt_map}\textit{b}) FWHM of 338~as, which is close to the Fourier-Transform-Limited (FTL) duration of 336~as. We have used Eq.~(\ref{eq:hh_sum}) to obtain the FTL pulse, with the assumption that all harmonics are perfectly phase-locked:
\[\phi_q = 0 \quad (or \quad constant) \quad \forall q\]
The high peak intensity achieved in our short pulses and the longer wavelength compared to typical Ti:Sapphire lasers are factors that reduce the magnitude of the attochirp \cite{Mairesse2003, Mairesse2004}. The residual nonlinear spectral phase contributes only minor amplitude modulations between pulses.

\subsection{Polarization gating}
\begin{figure}
    \centering
    \includegraphics[width=8cm]{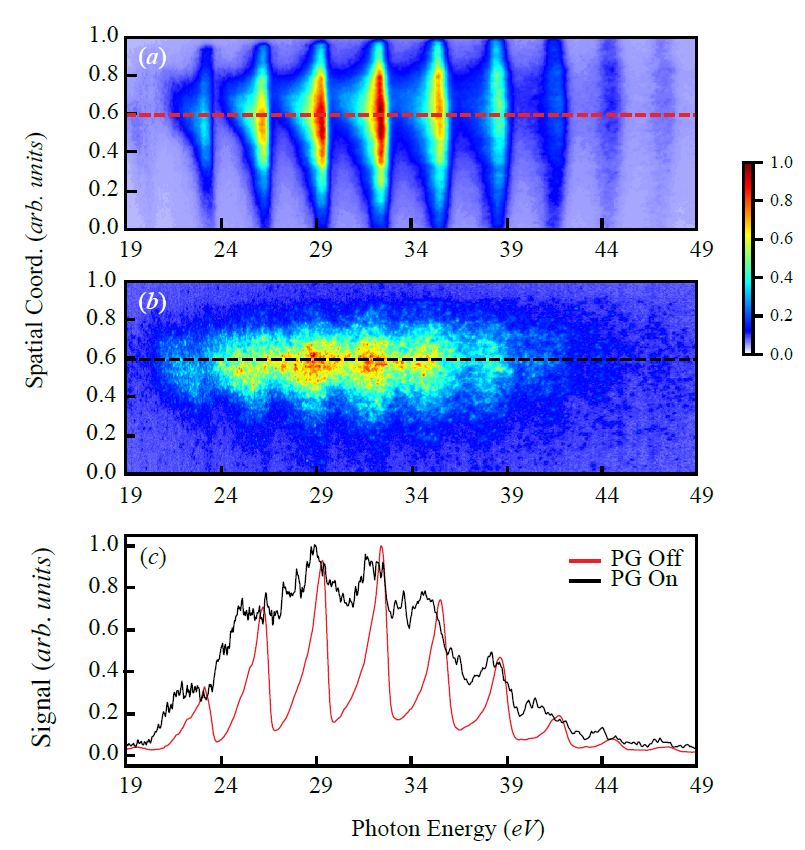}
    \caption{XUV spectra recorded (\textit{a}) in Argon gas without polarization gating, (\textit{b}) with polarization gating  and the respective linecuts are shown in (\textit{c}).}
    \label{fig_polarization_gating}
\end{figure}

Another key capability of the Compact beamline is the generation of isolated attosecond pulses. This is realized using a polarization gating (PG) setup as detailed in \cite{Kuhn2017,Tsafas2023}). The setup employs a series of waveplates to impose a temporally varying ellipticity on the driving IR pulse, such that high-harmonic generation occurs predominantly within a single half-cycle. For this purpose, the SYLOS laser was configured to deliver 11 fs pulses, while the waveplate arrangement was adjusted to yield an effective gate width of approximately half a laser cycle. Fig. \ref{fig_polarization_gating} displays single-shot XUV spectra recorded (\textit{a}) without and (\textit{b}) with PG applied to the SYLOS beam; corresponding on-axis lineouts are shown in (\textit{c}), clearly illustrating the transition from a comb-like to a quasi-continuous spectrum. Although a direct temporal characterization of the attosecond pulses was not performed, the emergence of a continuous spectrum serves as indirect evidence for the generation of near-isolated attosecond pulses \cite{Sola2006}. The residual comb-like modulations superimposed on the continuum further suggest that contributions from more than one optical cycle are present. Since the SYLOS laser was not carrier-envelope-phase (CEP) stabilized, when these measurements were done, subsequent shots produced variations in the harmonic spectra with this technique. To address this, an f-2f CEP tagging setup \cite{Tzallas2010} has been implemented on the beamline to enable post-processed data analysis.

\subsection{XUV pulse application}

The observation of a non-linear process initiated by an attosecond pulse train represents a crucial step toward exploiting the beamline for investigations of electron dynamics on the attosecond timescale. In particular, it demonstrates that the achieved peak intensity is sufficient to drive a well-defined two-photon response, thereby establishing the feasibility of studying higher-order processes under these conditions. On this basis, we have performed experiments on the two-photon double ionization of neon and argon. As the overall pulse duration is on the few femtosecond time scale, Ne$^{2+}$ is likely to be produced through a sequential process.  
The non-sequential process is expected to become dominant either for a pulse duration below 2 fs \cite{Orfanos2022} or for photon energies below the sequential threshold.

The experiment was carried out using XUV pulses generated in Argon with the SEA laser. Both the double gas jet target and the 30 cm gas cell we used provided a similar spectrum that peaks around 35 eV and extends from the cut-on of the aluminum filter at 18 eV to the cut-off of the silicon mirror around 50 eV. 
Selection of the bandwidth used in this experiment was accomplished with a multilayer mirror nominally centered at 40 eV \cite{Orfanos2022}.
The thus selected XUV spectrum includes the threshold for sequential generation of Ne$^{2+}$ located at 40.9 eV \cite{Orfanos2022, Sorokin2007} in the high energy wing. 
However, single photon double ionization at 61.5 eV \cite{Sorokin2007} is clearly excluded. 
We therefore expect the yield of Ne$^{2+}$ to follow a quadratic dependence on the photon flux. 
A common way to vary the photon flux while preserving all other pulse parameters is to vary the particle density of the generating medium \cite{Manschwetus2016, Nayak2018} which was also applied in the present case. 
A rough estimate of the peak intensity can be obtained from the measured  values for a pulse energy of 20 nJ before the multilayer mirror and an overall pulse duration of 4 fs \cite{Orfanos2022}. 
As the bandwidth of the mirror covers about 25$\%$ of the spectrum and the average reflectivity is 20$\%$, the pulse energy on target is around 1 nJ ($1.4 \times 10^8$ photons). With a conservative estimate of the focal spot size of 8~$\mu m$, the peak intensity could reach $0.9 \times 10^{12}$ W/cm$^2$ and a fluence of  $5 \times 10^{14}$ photons/cm$^2$. In this regime, the ratio of doubly to singly ionized neon is on the order of 2$\%$ as found from a solution of the rate equations for this process \cite{Orfanos2020}. The next higher ionization state Ne$^{3+}$ would require the absorption of four XUV photons where the step from Ne$^{2+}$ to Ne$^{3+}$ involves a direct two-photon absorption at the given photon energy \cite{Sorokin2007}. The absence of triply ionized states of neon is thus not surprising.

\begin{figure*}
    \centering
    \includegraphics[width= 14cm]{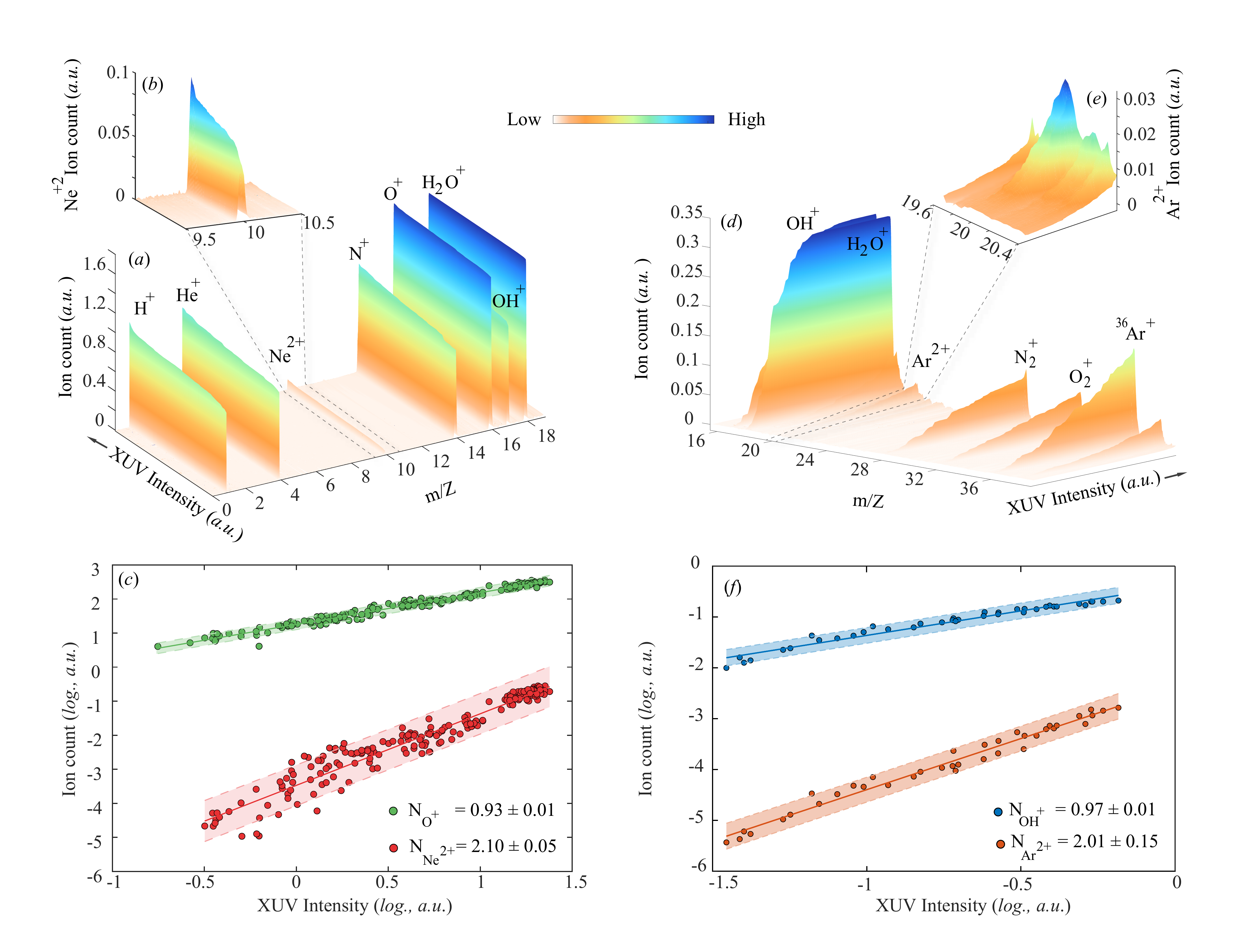}
    \caption {Two-photon ionization in Neon and Argon. The measurements were performed using Neon (Dual Ar gas jet at generation) and Argon (Dual Xe gas jet at generation) at the target region. (\textit{a, d}) Time-of-flight (ToF) ion mass spectra recorded as a function of XUV photon energy. (\textit{b, e}) Magnified views highlighting the nonlinear signals corresponding to Ne$^{2+}$ and Ar$^{2+}$, respectively. (\textit{c, f}) Dependence of the O$^+$ and Ne$^{2+}$ ion yields, as well as the OH$^+$ and Ar$^{2+}$ ion yields, on the XUV intensity. The intensity axes in (\textit{c, f}) are calibrated using the He$^+$ and N$_2^+$ signals, which scale linearly with the XUV intensity. Solid lines represent linear fits to the data, and the shaded regions indicate the 95\% prediction intervals for future observations.}
    \label{Ion_tof_yield}
\end{figure*}
Similar two-XUV-photon processes have also been observed in Ar at photon energies in the range of ~15-25 eV, generated in a double Xenon gas jet and transmitted by a 150 nm thick Sn filter. This photon-energy range lies well below the second ionization potential of Ar (43.4 eV), ensuring that double ionization via absorption of a single photon is energetically forbidden.

The yield of singly ionized species produced via single-photon ionization is proportional to the ionizing intensity, reflecting a linear process. Consequently, this yield can be used as a measure of the XUV intensity when examining the XUV intensity dependence of an interaction product.
As shown in the time-of-flight (ToF) mass spectra in Fig. \ref{Ion_tof_yield}(\textit{a, d}), several ion species originating from the background gas are observed. To elucidate the two-photon double ionization processes leading to the formation of Ne$^{2+}$ and Ar$^{2+}$, He$^+$ and N$_2^+$ were selected as reference ions, respectively.
At the selected signal magnification presented in Fig. \ref{Ion_tof_yield}(\textit{a, d}), the signals corresponding to singly ionized neon and argon exceed the detection threshold and are not shown here. The Ne$^{2+}$ and Ar$^{2+}$ peaks are clearly distinguishable at mass-to-charge ratios of 10 and 20, respectively.
Figures \ref{Ion_tof_yield}(\textit{c, f}) display the yields of Ne$^{2+}$ and Ar$^{2+}$, along with those of various other species, as functions of the He$^+$ and N$_2^+$ yields, and thus as functions of the XUV pulse intensity. On a log-log scale, the Ne$^{2+}$ and Ar$^{2+}$ ion yields exhibit slopes close to 2, whereas the other species show slopes around 1. This behaviour indicates that both Ne$^{2+}$ and Ar$^{2+}$ ions are generated through the absorption of two photons.

The observation of a nonlinear process initiated by an attosecond pulse train marks a crucial step toward utilizing the beamline for investigations of electron dynamics on the attosecond timescale through XUV-pump-XUV-probe approaches. In particular, it demonstrates that the achieved peak intensity is sufficient to drive a well-defined two-photon response, thereby confirming the feasibility of studying higher-order processes under these conditions. 
Furthermore, they highlight that nonlinear XUV processes can be efficiently induced by the beamline across different spectral regions.

\section{Conclusion}

In conclusion, we have successfully commissioned the SYLOS Compact GHHG 1 kHz beamline at ELI ALPS and demonstrated the generation of attosecond pulse trains at 1 kHz repetition rate, with an average duration of 336 as and XUV energies of 7.4 nJ on target (308 nJ at generation) across the 15–40 eV spectral range, achieved through quasi-phase matching using two gas jet sources. By applying polarization gating to the driving laser, we produced an XUV continuum, demonstrating the feasibility of generating isolated attosecond pulses.
The combination of loose focusing and high laser energy enabled high-flux harmonic generation and sequential double XUV ionization of neon and argon, paving the way for real-time observation of electron dynamics on the attosecond timescale without a dressing IR field.
These results highlight the exceptional capabilities of the ELI ALPS beamline, offering open-access, user-oriented infrastructure for ultrafast studies in atomic, molecular, and condensed matter systems. By delivering attosecond pulses at high repetition rates and high flux, the facility empowers experiments that were previously inaccessible, including investigations of higher-order non-linear processes, complex electron correlations, and new regimes of attosecond science. This beamline at ELI ALPS establishes a versatile and powerful platform for advancing both fundamental and applied research at the forefront of ultrafast XUV science.

\section*{Acknowledgments}

\subsection*{General} 
We thank Tibor Nagy and M\'aty\'as Mozs\'ar for their help in preparing realistic beamline schematics. We also acknowledge Sergei K\"uhn for his contribution towards the beamline construction, management, development, data analysis and scientific discussions. Artificial intelligence was used as aid in text editing.

\subsection*{Author Contributions} 
A.N. and M.D. constructed and commissioned the beamline and performed the experiments. J.C., Sz.T. and P.G. operated and maintained the laser system and assisted in beam characterization. I.O., E.S., P.T., D.C. contributed to the realization of the experiments and data interpretation. S.M., D.R., and N.A. performed data analysis, visualization, and interpretation of the results. Z.D. supervised the preparation of the manuscript. All authors contributed to manuscript preparation.

\subsection*{Funding}
ELI ALPS is supported by the European Union and co-financed by the European Regional Development Fund (ERDF) (GINOP-2.3.6-15-2015-00001).

\subsection*{Competing interests}
The authors declare that they have no competing interests.

\subsection*{Data Availability}
The data that support the findings of this study are available from the authors upon request.

\printbibliography

\end{document}